\documentclass[10pt, a4paper]{article}

\usepackage[final]{lrec2026} 
\usepackage{multirow}
\usepackage{multicol}
\usepackage[most]{tcolorbox}
\usepackage{setspace}
\usepackage{caption}

\title{Report-based Recommendations for Policy Making \\ and Agency Operations: Dataset and LLM Evaluation}

\name{Aleksandra Edwards, Thomas Edwards, Jose Camacho-Collados, Alun Preece} 

\address{Cardiff University \\
         Senghennydd Rd, Cardiff CF24 4AG \\
         \{EdwardsAI, EdwardsTJ1, CamachoColladosJ, PreeceAD\}@cardiff.ac.uk\\}

\abstract{
 Large Language Models (LLMs) are extensively used in text generation tasks. These generative capabilities bring us to a point where LLMs could potentially provide useful insights in policy making or agency operations. In this paper, we introduce a new task consisting of generating recommendations which can be used to inform future actions and improvements of agencies work within private and public organisations. In particular, we present the first benchmark and coherent evaluation for developing recommendation systems to inform organisation policies. This task is clearly different from usual product or user recommendation systems, but rather aims at providing a basis to suggest policy improvements based on the conclusions drawn from reports. Our results demonstrate that state-of-the-art LLMs have the potential to emphasize and reflect on key issues and learning points within generated recommendations. 
 \\ \newline  }

\begin{document}

\maketitleabstract

\section{Introduction}\label{intro}
Recent LLMs~\cite{brown2020language, touvron2023llama2,dubey2024llama,chowdhery2023palm} have shown exceptional abilities in text generation tasks such as summarisation~\cite{zhang2024benchmarking,xie2023survey} and story generation ~\cite{tang2022context, razumovskaia-etal-2024-little-red}, among others, achieving results comparable to human-created text. Given the ability of LLMs to understand instructions written in natural language (\textit{`prompts'}), the majority of work is focused on utilising prompt-based 
approaches for adapting pre-trained models to different domains and tasks~\cite{viswanathan2023prompt2model,plaza2023leveraging}. As LLMs continue to scale, research has increasingly focused on their potential for more specialized applications that have traditionally relied on domain experts~\cite{huang2024good}. One such example is Court View Generation (CVG) in the legal domain~\cite{li2024enhancing,yue2021circumstances,wu2023precedent}, where the goal is to generate textual interpretations of judgment results. Progress in this space has primarily relied on integrating domain-specific knowledge into pre-trained LLMs, which has proven more effective than applying generic models alone~\cite{li2024enhancing,wu2023precedent,yue2021circumstances}. This highlights the need for methods that combine the reasoning capabilities of LLMs with expert insights to tackle specialised text generation challenges. However, research into such domain-specific generative tasks remains limited, with most work concentrated within the broader field of Legal AI. Moreover, while discussions around LLM governance and risk assessment are gaining traction~\cite{goanta2023regulation}, initiatives such as RegNLP remain focused on regulatory alignment rather than develop models for domain-specific generative tasks.
\begin{figure}[t!]
		    \begin{center}
		     \includegraphics[scale = 0.125]{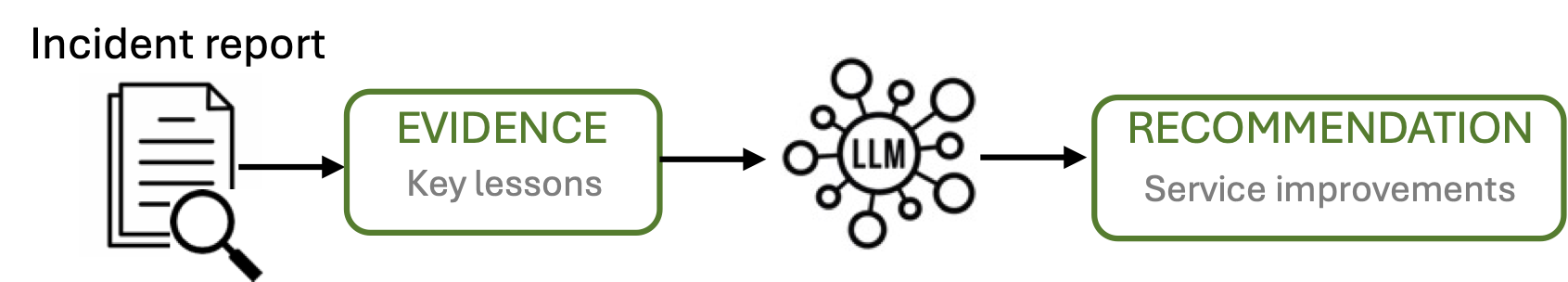}
		
        \caption{An overview of the recommendation generation pipeline.}\label{example}
       
	  \end{center}
    \end{figure}
We present the task of recommendation generation, a new frontier for natural language generation (NLG) where the goal is to help practitioners in the public sector write actionable recommendations that inform policy-making and improve service delivery. Unlike standard text generation tasks (e.g., narrative continuation, product recommendation), recommendation generation operates in a highly dynamic environment: requirements evolve rapidly across government and service agencies, the language used is diverse and specialised, and outputs must balance domain-specific accuracy with clarity and practical applicability. This makes the task both technically challenging and socially impactful, with direct implications for improving support to vulnerable populations.

Our main contributions are as follows:

\textbf{(1) A New Task on Policy-focused Recommendation Generation:} We introduce a new NLG task that investigates how LLMs can assist public sector practitioners in drafting recommendations to support policy-making processes for improving service delivery to vulnerable individuals. \\ \textbf{(2) PubRec-Bench Benchmark Dataset:} We release a unified benchmark dataset collected from three well-established, independent sources, covering both UK- and US-based contexts sources: the \textit{UK Care Homes} reports on care quality for vulnerable adults, the \textit{US Children’s Bureau} reports on foster care and adoption services, and the \textit{NSPCC} reports on serious incidents involving children.  \\\textbf{(3) Evaluation of LLM Performance and Metrics:} We conduct extensive evaluation of three state-of-the-art LLMs for recommendation generation, using similarity-based metrics, LLM-based evaluation, and human assessment. Results highlight both the promise of LLMs for this task and the discrepancies across evaluation methods, underscoring the need for tailored evaluation approaches.

\section{Related Work}\label{relwork}
NLG aims to produce text from a given input data where the generated output needs to satisfy certain language properties and task requirements~\cite{tang2022context}. The enhancements in the field in the recent years in terms of creating more powerful language models, have lead to an increased research into how to utilise these tools for more challenging problems and domains requiring subject matter expertise or/and lack training data. 
Many approaches tackling the data sparsity problem rely on prompting (in-context learning) methods for generating text. Prompting is a technique which allows to guide LLMs into performing downstream tasks by providing either instructions written in natural language (zero-shot) or providing a few examples (few-shot)~\citetlanguageresource{razumovskaia-etal-2024-little-red}. Existing work has shown that prompting can lead to a strong performance in various tasks such as question answering~\citetlanguageresource{chowdhery2023palm,agrawal2023qameleon} and open-ended natural language generation~\cite{tang2022context}, even in some cases to comparable or even better performance than standard fine-tuning techniques especially in the absence of training corpora~\cite{gao2021making, mosbach2023few}.

Research into utilising LLMs for text generation in more specialised domains is mainly focused on summarisation tasks for the clinical and law domains. For instance, in the medical domain there is an increased work on developing summarisation tools to support clinical information retrieval and management~\cite{xie2023survey,xie2023knowledge,lopez2024evaluation}. 
In the legal domain, there has been an increased interest in developing LLM-driven approaches for court view generation (CVG)~\cite{li2024enhancing,yue2021circumstances,yu2022legal,wu2023precedent, tyss2024lexabsumm}. CVG is a natural language generation (NLG) task, which aims to generate court views based on the plaintiff claims and the fact descriptions related to a given court case~\cite{li2024enhancing}. The majority of research in the area is focused on incorporating domain knowledge and LLMs for the task~\cite{wu2023precedent,li2024enhancing,yue2021circumstances} where results show the need for more domain-targeted approaches when it comes to highly specialised texts. For instance, the approach proposed by \newcite{li2024enhancing} is based on injecting claim-related knowledge such as keywords and label definitions within the prompt encoder of the model. The authors of~\cite{wu2023precedent} propose a framework that incorporates pre-trained LLMs, prompting techniques and small domain-trained language models. A work by~\cite{savelka2023can} takes a different approach where the authors evaluate the capability of GPT4 for court opinions to interpret legal concepts. The work showed that GPT-4, guided only by in-context learning techniques, can give similar performance to a well-trained law student annotators. This work highlights interesting research avenues for exploring text generation capabilities of LLMs in more specialised domains. However, prior work is mainly focused on the LegalAI domain. Further, there is a growing concern about suitability of existing evaluation measures when it comes to text generation~\cite{liusie-etal-2024-llm,panickssery2024llm,gao2024llm,khashabi2021genie,chaganty-etal-2018-price}, especially within more high risk domains and tasks~\cite{lopez2024evaluation}. However, the aforementioned research lack discussion on suitability of evaluation metrics used. In this work, we introduce a new task and dataset for report-based recommendation generation in policy and agency contexts, and provide a thorough evaluation with critical analysis of the suitability of existing NLG evaluation metrics.

\section{PubRec-Bench: Recommendation Generation Benchmark}\label{pubrec_bench}
In this section, we describe the task of recommendation creation for informing policy making in the public sector (Section~\ref{taskdesc}), the process of collecting and unifying relevant datasets (Section~\ref{datasetcol}), and their statistics (Section~\ref{datastats}).
\begin{table*}[hbt!]
        \centering
        \resizebox{\linewidth}{!}{
        \begin{tabular}{|l|p{8.15cm}|p{7.05cm}|}\hline
           \textbf{Dataset}  &\textbf{Evidence}& \textbf{Recommendation}\\\hline
            \textbf{UK Care Homes}& The social care and wellbeing learning and development team action planning
framework was substantial but it was not possible to evaluate the impact.
The family and community support action plan 2012 was a draft and had not been
fully populated... &   The social work services should ensure that annual reviews of people placed in care homes are carried out by clarifying the appropriate responsibilities and timescales.  \\\hline
            \textbf{US Children Bureau}& Use of the supplemental issuance code as a `catch-all' for certain costs. Regional
Office staff were required to manually review and request additional information in 26 cases in
order to determine the purposes for the supplemental issuances and whether they were for
allowable title IV-E maintenance expenditures...& The state should provide guidance to counties to be sure that it is able to segregate out the reasons why the supplemental issuance code is used so that the various types of
supplemental payments may be identified. \\\hline
            \textbf{NSPCC reports}&The work would have benefitted from exploration of key relationships and extended family
on both sides.... A genogram would have enabled further exploration
of the nuances of the family. Whilst it is unlikely that this would not have impacted on the
outcome, it would have provided a more complete picture...  &SHIELD to develop a 7 minute briefing and top tips for practitioners about how to act on gut feelings and professional curiosity. A task and finish group should lead on this work
which should include refreshing and promoting SHIELD’s website content.\\\hline
        \end{tabular}
        }
        \caption{Examples of extracted evidence and  recommendation pairs per dataset type.}
        \label{examplesrecom}
    \end{table*}

    \subsection{Task Description}\label{taskdesc}
    Local authorities and community safety partnerships often need to produce reports in order to reflect on public services or identify and describe related events that precede a serious incident, for example involving a child or vulnerable adult. A key role of these documents is to reflect on agencies' roles and the application of current practices in social care provision and crime prevention. These reports, despite being quite diverse in structure and topics, need to contain key lessons learned (\textbf{\textit{evidence)}} of good or bad practices that are used to derive a set of (\textbf{\textit{recommendations)}}. These recommendations are disseminated (independent of the reports) across relevant institutions in order to inform the development of policy making for improving service delivery across different governmental sectors. The development of these recommendations can be biased and a resource- consuming task, resulting very often in the creation of bad quality content. In this paper, we explore if and how LLMs can be used to support practitioners in writing high quality recommendations. Specifically, given an evidence of lessons learned, our task consists of generating a recommendation which reflects on and it is consistent with the provided information.

    \subsection{Dataset Collection and Unification}\label{datasetcol}
        
   We collected three datasets, consisting of reports reviewing agencies work related to the provision of services to vulnerable individuals. These reports are lengthy and contain information irrelevant to the recommendation generation task, such as information regarding the reviewing board, incident description and timeline of events. Thus, for the purposes of our analysis, we have extracted the evidence from the reports as these contain sufficient information for generating recommendations, and this setting can help prevent possible LLM hallucinations with irrelevant information from the reports. Further, the reports have very diverse structure and content within and across the different data sources making it hard to identify evidence with associated recommendations. Each document was individually reviewed by two human annotators, who extracted evidence and corresponding recommendations only when there was a clear, justifiable link between the two. Sections with recommendations lacking supporting evidence were excluded, ensuring that only high-confidence evidence–recommendation pairs were retained. This thorough, manual curation process was performed to minimise noise and enhance the reliability of our evaluations. While resource-intensive, we believe that this rigorous approach establishes a robust foundation for future scaling and broader dataset expansion. All reports are publicly available to download via their websites.  Examples of evidence and recommendation pairs for each dataset are given in Table~\ref{examplesrecom}. The three datasets have been sourced from large, authoritative repositories that document service provision in highly sensitive domains such as child protection and adult care. Importantly, these reports are produced only under specific institutional circumstances (e.g., inspections or serious case reviews), which explains the relatively limited volume of available data. Despite this, the reports span multiple policy domains and two different national systems (UK and US), offering a diverse collection of evidence–recommendation pairs that are highly relevant to the task of recommendation generation. 
    The datasets are available at: \url{https://github.com/AleksEdwards/PubRec_Data} \\ 
    \noindent\textbf{UK Care Homes reports.} The \textit{`UK Care Homes'}~\footnote{UK Care Insp: \url{www.careinspectorate.com}} dataset consists of reports produced by The Care Inspectorate in order to reflect on the quality of care homes for vulnerable adults in UK. The website contains roughly around 300 reports, however, not all of them contain recommendations. In order to allow comparison between generated and human-written recommendations we have excluded reports with missing recommendations from our collection.\\
    \noindent\textbf{US Children's Bureau reports.} The \textit{US Children's Bureau} dataset~\footnote{Children's Bureau: \url{https://acf.hhs.gov/cb}} consists of reports that assess the quality of foster care and adoption services in the US. Children's Bureau is an agency within the Administration for Children and Families, which is part of the U.S. Department of Health and Human Services. The learning points and recommendations from the reports are used to help prevent child abuse and neglect, create better adoption services and foster care. \\ 
    \noindent\textbf{NSPCC reports.} NSPCC (The National Society for the Prevention of Cruelty to Children) is UK's leading children’s charity that specialises in child protection and prevention of child abuse. The NSPCC reports\footnote{NSPCC: \url{https://library.nspcc.org.uk}} consists of case reviews written by UK-based Local Safeguarding Children Boards (LSCBs).

\subsection{Data Statistics}\label{datastats}
The three datasets consist of 110 reports and 493 recommendations in total (see Table~\ref{tab:datastats}). Considering that these reviews are produced only when a serious incident  occurs, our collection represents a substantial subset of the total number of reports available. Further, reports for all datasets have an average length above 7,000 tokens (see Table~\ref{tab:datastats}) which makes processing in their entirety a challenging task, which could be a subject to future research. 
\begin{table}[hbt!]

	\centering
	\resizebox{\linewidth}{!}{
	\begin{tabular}{|l||c|c|c|}\hline
 &\textbf{UK Care}&\textbf{US Children}&\textbf{NSPCC}\\\hline\hline
 \textbf{\# reports}&22&48&40\\\hline
 \textbf{\# \textit{recs}}&94&122&276\\\hline
 \textbf{Avg \# \textit{recs} per report}&4&2&7\\\hline
 \textbf{Avg \# tokens per \textit{recs}}&34&118&61\\\hline
 \textbf{Avg \# tokens per evidence}&742&254&219\\\hline
 \textbf{Avg \# tokens per reports}&9,567&7,943&13,120\\\hline

	\end{tabular}
	}
	\caption{Dataset statistics where \textit{`\#reports'} refers to number of reports per dataset, \textit{`\#recs'} refers to number of recommendations per dataset, \textit{`avg'} refers to average.}\label{tab:datastats}
	\end{table}
    
\section{Experimental Setting} \label{sec:settings}
    \subsection{Recommendation Generation}\label{recgen}
    The aim of the paper is to analyse the feasibility of incorporating LLMs within the process of writing recommendations for improving public services and agencies work based on evidence collected from previous good and bad practices. We would like to note that we focus on evaluating models which are known to provide state-of-the-art performance for text generation tasks, especially in low-resource settings. Therefore, performing extensive evaluation of a large variety of different models is outside the scope of the paper.
    
    \noindent\textbf{Comparison Models.}\label{compmodels} For the purposes of our analysis, we compare three different models. These are: \textbf{(1) OpenAI GPT-4o model} which is one of the most advanced models released within the NLP space and it is well known for its impressive zero- and few-shot capabilities~\cite{savelka2023can,brown2020language}. \textbf{(2) Command R+} is Cohere's most powerful and newest large language model, optimized for conversations and long-context tasks and it consists of 104B parameters.  \textbf{LLaMa 3 model} which is known to be one of the most advanced open source language models~\citetlanguageresource{dubey2024llama}. We use LLaMA 3 model with 8 billion parameters, pre-trained with instructions, downloaded from HuggingFace~\citetlanguageresource{wolf2019huggingface}\footnote{Parameters are available in the Appendix.}. \\

    \noindent\textbf{Prompting.}\label{icl} Given the limited amount of annotated data, we use the in-context learning method to generate recommendations. As described in Section~\ref{relwork}, prompting can lead to better results compared to fine-tuning techniques when data is limited. Importantly, prompt-based approaches also mirror realistic deployment scenarios, where practitioners may have limited data but access to powerful general-purpose LLMs. Moreover, fine-tuning in high-stakes policy contexts can raise data privacy concerns and typically requires significantly larger annotated datasets, which are not yet feasible in this domain. In addition, our objective is to analyze the extent to which state-of-the-art LLMs can perform complex tasks with limited resources. We generate recommendations using prompting in zero-shot and one-shot settings, where the model is given a description of the task and supporting evidence. We conduct experiments using two prompts: a generic prompt and a domain-expert-developed prompt, to assess how the amount of information in the prompt affects model performance. For the creation of \textit{`Prompt 1'}, we followed examples provided by OpenAI and Meta. We also followed the design principles described in \newcite{reynolds2021prompt} to create self-explanatory prompts that are intuitive and easy to use from the user's perspective. To create \textit{`Prompt 2'}, we asked subject matter experts (see Section~\ref{humeval_desc} to design the prompt. 

   \begin{tcolorbox}[colback=blue!5!white,colframe=blue!25!black,title= Prompt 1 for generating recommendations]
        Provide a recommendation for improving agencies work and services related to children care and children services. The recommendation should reflect on the information given in the report:\\
        Evidence:[Evidence]
        \end{tcolorbox}
        
        \begin{tcolorbox}[colback=blue!5!white,colframe=blue!25!black,title= Prompt 2 for generating recommendations]
        Based on a summary of evidence from this report, generate a concise recommendation with particular focus on what would improve or resolve the issues raised within the information. Please do not include context or rationale at this stage:\\
        Evidence:[Evidence]
        \end{tcolorbox}    

    \subsection{Evaluation}\label{evaluationoverall}
        We evaluated the generated recommendations using three types of evaluation measures, ie., similarity metrics, LLM-based evaluation, and human-based evaluation. This allows us to capture different aspects of how well the models perform for recommendation generation as well as to allow analysis into the suitability of these measures for evaluating NLG tasks.
        
       \noindent\textbf{Similarity Metrics.} We use traditional reference-based evaluation metrics such as BLEU~\cite{papineni2002bleu} and ROUGE~\cite{lin2004rouge} which measure the extent to which generated content matches the n-grams of the reference text. In particular, we use ROUGE-L to measure the longest common subsequence (LCS). In addition, we use BERTScore~\cite{zhang2019bertscore}, an embedding-based method which uses embedding representations of the reference and the target text to compute semantic similarity between them. This metric could be better suited to the varying size of recommendations. Nonetheless, we anticipate that these automatic metrics may have shortcoming when it comes to the evaluation and therefore, we propose both an additional automatic LLM-based metric and a human evaluation.\\
        
        \noindent\textbf{LLM-based Evaluation.} We use a prompt-based approach~\cite{gao2024llm} and measure the factual alignment between the reference and targeted recommendations using each one of the language models. The prompt is created following the same principles used for recommendation generation in Section~\ref{recgen}. Within the prompt, we specify the evaluation criteria based on a 3-point Likert scale where 1 refers to the lack of any factual alignment between the recommendations and 3 refers to a complete factual alignment between them. We use the same scale for the human evaluation to allow comparison between the evaluation approaches.

      \begin{tcolorbox}[colback=red!5!white,colframe=red!75!black,title=Prompt for evaluating recommendations]
        You are given two recommendations (Recommendation 1 and Recommendation 2). Your task is to measure the factual alignment between the two recommendations using a scale from 1 to 3 where 1 refers to the lack of any factual alignment between the recommendations and 3 refers to a complete factual alignment between them. \\Evaluation Form: Answer by starting with 'Rating:' and then give the explanation of the rating on the next line by 'Rationale:
        \end{tcolorbox}
        
        \noindent\textbf{Human Evaluation.}\label{humeval_desc} During evaluation, participants are given the generated recommendation, the evidence used to generate the recommendation, and the human-created recommendation. Each recommendation is evaluated by five subject matter experts using a 3-point Likert scale where 1 is worst and 3 is best. Finally, considering the highly specialised nature of the datasets which require domain experts for evaluation, we performed these experiments for 240 randomly selected recommendations across the three datasets. The subject matter experts were selected through an interview process and all have experience in dealing with policymaking processes for governmental institutions. For conducting human evaluation\footnote{The eval. sheet is available in the Appendix.}, we followed principles described in previous work \cite{chhun2022human,li2024enhancing}. We outlined 5 main criteria for conducting the evaluation: \textbf{(1) Fluency} --- measures the quality of the text including grammatical errors and repetitions; \textbf{(2) Coherence} --- measures whether the recommendation makes logical sense.   \textbf{ (3) Relevance to the evidence} --- measures whether the recommendation is meaningful given the evidence;  \textbf{(4) Relevance to the human-created recommendation} --- measures the factual alignment between the two recommendations (we use the same criteria for LLM-based evaluation to allow comparison between the two measures); \textbf{(5) Is the recommendation `Actionable'? (yes/no)} --- shows if the recommendation has practical application and could be implemented as part of a policy.
 
\section{Results and Analysis}\label{resandanalysis}
 The aim of our analysis is to (1) identify to what extend state-of-the-art LLMs can perform recommendation generation for informing policy making, as well as (2) analyse the suitability of existing evaluation metrics for the task.

 \subsection{Automatic Evaluation}\label{automaticeval}

A comparison between the performance of the generation models for the two prompts (see Table~\ref{tab:prompts}) showed consistently higher results for prompt 2 (i.e., the prompt designed by subject matter experts). This shows the importance and need to involve domain expertise not only during the evaluation process of LLM-based approaches but also during the development of the LLM-based system. 

\begin{table*}[ht]
	       \centering
        \resizebox{\linewidth}{!}{
      
        \begin{tabular}{|l|l||c|c|c|c|c|c|c|}\hline
        \textbf{Data}&\textbf{prompt}&\textbf{BERT-Score (F1)}&\textbf{ROUGE-L (F1)}&\textbf{BLEU Score}&\textbf{GPT-based eval.}&\textbf{LLaMA-based eval.}&\textbf{Cohere-based eval.}\\\hline\hline
        \multirow{2}{*}{UK Care Homes}&prompt 1&0.446&0.107&0.004&1.953&1.719&1.939\\\cline{2-8}
        &prompt 2&0.555&0.189&0.008&2.168&1.806&2.048\\\hline\hline
        \multirow{2}{*}{US Children’s Bureau}&prompt 1&0.466&0.134&0.011&2.519&2.019&2.067\\\cline{2-8}
        &prompt 2&0.584&0.231&0.016&2.570&1.997&2.056\\\hline\hline
         \multirow{2}{*}{NSPCC reports}&prompt 1&0.445&0.108&0.007&2.197&1.904&1.949\\\cline{2-8}
         &prompt 2&0.557&0.189&0.023&2.218&1.902&2.029\\\hline

         \end{tabular}
         }
            \caption{Averaged evaluation results across all LLMs for generating recommendations using prompt 1 and prompt 2 (prompts described in Section \ref{sec:settings}) in the zero-shot setting. The evaluations are based on similarity metrics (`BERT Score', `ROUGE-L', `BLEU Score') and LLM-based evaluations using GPT (`GPT-based eval.'), LLaMA (`LLaMA-based eval.'), and Cohere (`Cohere-based eval.'). }\label{tab:prompts}
	\end{table*}

Table~\ref{tab:evalresfull_automated} shows individual model results of recommendation generation based on automatic metrics. The similarity metrics, especially BLEU Score and ROUGE-L show quite low results across datasets, settings and prompts, and models in comparison to LLM-based evaluation. This highlights the limitations of these traditional automatic metrics to capture the factual correctness of generated text as well as semantic similarities for more complex NLG tasks. In contrast, LLM-based evaluation (regardless of model used) shows a good quality of generated recommendations regarding factual consistency with the gold standard. Specifically, the average score for each LLM-based evaluation, regardless of the model used to generate recommendations, varies between 1.7 and 2.5. The results suggest a slightly better performance for GPT4-o and thus we use recommendations generated with this model to perform human evaluation. Overall, evaluation results show a better performance in the US Children's Bureau dataset, which can be attributed to the fact that the `evidence' for these documents are shorter passages in comparison to the UK Care Home or the NSPCC dataset. Another potential reason is the regional differences between the datasets where the US-based reports cover a larger and potentially better represented location within the training set of these models. 

\noindent \textbf{Zero-shot vs. one-shot.} An important observation is that models consistently perform better in the zero-shot setting compared to the one-shot setting. One possible reason for this is the high variability in the evidence and recommendation formats, which suggests that traditional in-context learning approaches relying on a small number of labeled examples may be insufficient for improving model performance in this domain. Instead, more dynamic and domain-specific adaptation strategies may be needed to effectively guide the models. 

\begin{figure*}[hbt!]
		    \begin{center}
            \includegraphics[scale=0.11]{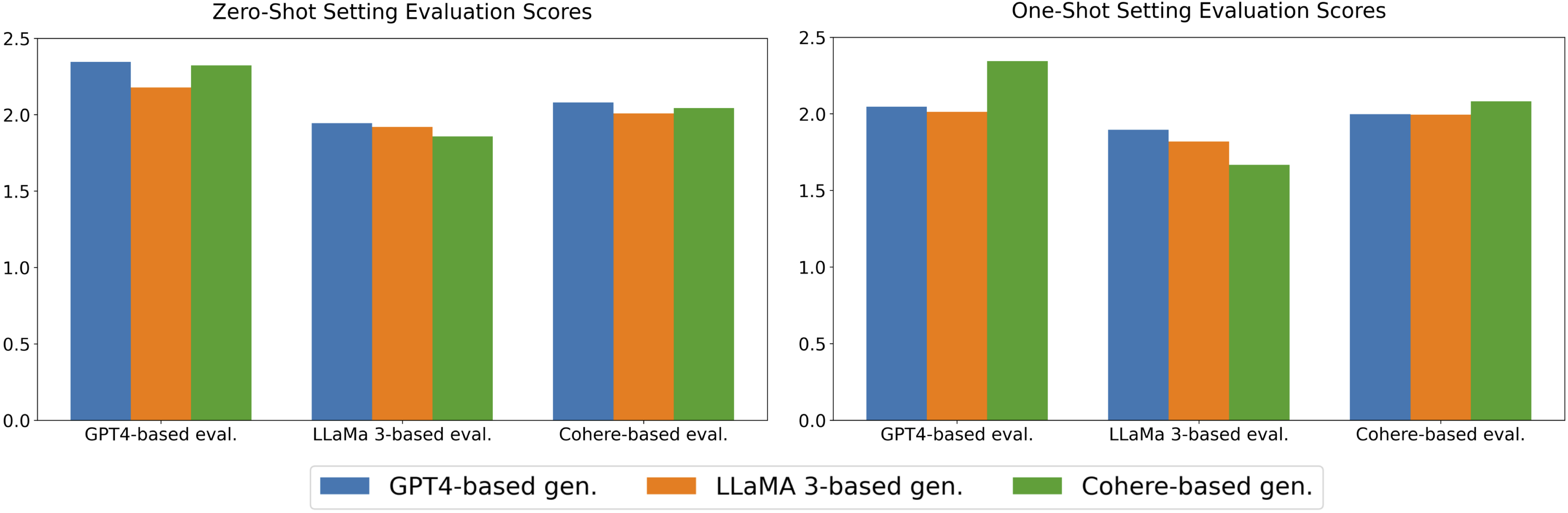}
        \caption{Comparison of LLM-based evaluations (\textit{`eval'}) in zero-shot settings (left) and one-shot settings (right) for recommendations generated by each model across the three datasets.}\label{biaseval}
       
		    \end{center}
    \end{figure*}

\noindent \textbf{LLM evaluators.} A comparison of the three LLM-based evaluation models for zero- and one- shot setting (see Figure~\ref{biaseval}), where the scores are averaged across the three datasets, shows that the GPT4 and Cohere-based models give a higher score to their own outputs. This suggests a potential bias for these models towards their own generations \cite{kocmi2023large} which shows the need for further research into how best to utilise these models for evaluation tasks. 

\renewcommand{\arraystretch}{1.05}
\begin{table*}[ht]
	       \centering
        \resizebox{\linewidth}{!}{
      
        \begin{tabular}{|l|l|l||c|c|c|c|c|c|c|}\hline
        \textbf{Data}&\textbf{Setting}&\textbf{Gen Model}&\textbf{Bert-Score (F1)}&\textbf{Rouge-L (F1)}&\textbf{BLEU Score}&\textbf{GPT-based eval}&\textbf{LLaMA-based eval}&\textbf{Cohere-based eval}\\\hline\hline
        \multirow{6}{*}{UK Care Homes}&zero&GPT 4-o&\textbf{0.569}&0.181&0.010&\textbf{2.183}&\textbf{1.903}&\textbf{2.043}\\\cline{2-9}
        &zero&Cohere&0.552&0.171&0.005&2.140&1.720&2.000\\\cline{2-9}
        &zero&LLaMA&0.545&\textbf{0.189}&\textbf{0.009}&2.182&1.795&2.102\\\cline{2-9}
        &\multicolumn{2}{|l||}{\textit{AVERAGE zero-shot}}&\textit{0.555}&\textit{0.189}&\textit{0.008}&\textit{2.168}&\textit{1.806}&\textit{2.048}\\\cline{2-9}
        &one&GPT 4-o&0.573&0.183&0.011&2.086&1.860&\textbf{2.075}\\\cline{2-9}
        &one&Cohere&\textbf{0.578}&0.189&0.006&\textbf{2.237}&\textbf{1.913}&2.081\\\cline{2-9}
        &one&LLaMA&0.542&\textbf{0.190}&\textbf{0.018}&1.806&1.667&1.978\\\cline{2-9}
        &\multicolumn{2}{|l||}{\textit{AVERAGE one-shot}}&\textit{0.564}&\textit{0.190}&\textit{0.012}&\textit{2.043}&\textit{1.813}&\textit{2.045}\\\hline\hline
        \multirow{6}{*}{US Children’s Bureau}&zero&GPT 4-o&0.583&0.224&0.013&2.594&\textbf{2.009}&\textbf{2.113}\\\cline{2-9}
        &zero&Cohere&\textbf{0.594}&\textbf{0.246}&\textbf{0.020}&\textbf{2.612}&1.991&2.095\\\cline{2-9}
        &zero&LLaMA&0.575&0.222&0.014&2.504&1.991&1.959\\\cline{2-9}
        &\multicolumn{2}{|l||}{\textit{AVERAGE zero-shot}}&\textit{0.584 }&\textit{0.231}&\textit{0.016}&\textit{2.570}&\textit{1.997}&\textit{2.056}\\\cline{2-9}
        &one&GPT 4-o&0.572&0.221&0.015&\textbf{2.273}&1.942&1.917\\\cline{2-9}
        &one&Cohere&\textbf{0.588}&\textbf{0.243}&\textbf{0.017}&2.645&\textbf{2.017}&\textbf{2.132}\\\cline{2-9}
        &one&LLaMA&0.547&0.202&0.008&2.058&1.909&1.974\\\cline{2-9}
        &\multicolumn{2}{|l||}{\textit{AVERAGE one-shot}}&\textit{0.569}&\textit{0.222}&\textit{0.013}&\textit{2.325}&\textit{1.956}&\textit{2.008}\\\hline\hline
         \multirow{6}{*}{NSPCC reports}&zero&GPT 4-o&\textbf{0.567}&0.188&0.026&\textbf{2.258}&\textbf{1.920}&\textbf{2.084}\\\cline{2-9}
         &zero&Cohere&0.550&0.179&0.015&2.218&1.865&2.036\\\cline{2-9}
         &zero&LLaMA&0.554&\textbf{0.202}&\textbf{0.028}&2.178&1.910&1.967\\\cline{2-9}\cline{2-9}\cline{2-9}
        &\multicolumn{2}{|l||}{\textit{AVERAGE zero-shot}}&\textit{0.557}&\textit{0.189 }&\textit{0.023}&\textit{2.218}&\textit{1.902}&\textit{2.029}\\\cline{2-9}
         &one&GPT 4-o&0.555&0.173&0.013&2.047&1.884&2.000\\\cline{2-9}
         &one&Cohere&\textbf{0.561}&0.179&0.014&2.149&\textbf{1.898}&\textbf{2.034}\\\cline{2-9}
         &one&LLaMA&0.560&\textbf{0.188}&\textbf{0.028}&\textbf{2.175}&1.880&2.031\\\cline{2-9}
         &\multicolumn{2}{|l||}{\textit{AVERAGE one-shot}}&\textit{0.558 }&\textit{0.180}&\textit{0.018}&\textit{2.123}&\textit{1.887}&\textit{2.023}\\\hline

         \end{tabular}
         }
            \caption{Complete evaluation results by dataset and generation model, based on similarity metrics (`BERTScore', `ROUGE-L', `BLEU Score') and LLM-based evaluations using GPT (`GPT-based eval'), LLaMA (`LLaMA-based eval'), and Cohere (`Cohere-based eval'). All generations were produced using Prompt 2.}\label{tab:evalresfull_automated}
	\end{table*}

    \subsection{Human Evaluation}\label{reshumaneval}

Table~\ref{humanevalReszero} show a good overall performance of GPT-4o for recommendation generation across the three datasets where the average score across the majority of criteria is above 2.5. Similarly to the automatic evaluation, generation models are shown to perform better in zero-shot rather than one- shot by the human evaluation as well. These results also show higher overall score for the `relevance to the evidence'-based criteria versus `relevance to the human-created recommendation' (0.5 difference in score). This suggests that a strength of LLMs in NLG is in providing a different perspective for the task/input which can be useful to users, versus simply recreating the human gold standard. This also highlights the need for more task-targeted and purpose-oriented evaluation metrics.  In addition, the results for criterion \textit{(5) `Is the recommendation actionable?' } are promising (see Table~\ref{actionable}). In the zero-shot setting, annotators agreed that around 60\% of the recommendations across all datasets were actionable. In the one-shot setting, the figure was around 57\%. These findings suggest that the generated recommendations have meaningful practical value and potential applicability within policymaking processes.

 \begin{table}[hbt!]
	       \centering
        \resizebox{\linewidth}{!}{
         \begin{tabular}{|l|l||c|c|c|}\hline
         &\textbf{setting}&\textbf{UK Care}&\textbf{US Children}&\textbf{NSPCC}\\\hline
        \textbf{Fluency}&zero&2.753&2.877&2.787\\\hline
         \textbf{Coherence}&zero&2.887&2.970&2.890\\\hline
         \textbf{Rel. to the evidence}&zero&2.790&2.863&2.850\\\hline
         \textbf{Rel. to human rec.}&zero&2.553&2.537&2.463\\\hline\hline
         \textbf{AVERAGE}&\textbf{zero}&\textbf{2.746}&\textbf{2.811}&\textbf{2.748}\\\hline\hline
         \textbf{Fluency}&one&2.467&2.603&2.653\\\hline
         \textbf{Coherence}&one&2.767&2.787&2.837\\\hline
         \textbf{Rel. to the evidence}&one&2.740&2.753&2.700\\\hline
         \textbf{Rel. to human rec.}&one&2.307&2.437&2.277\\\hline\hline
         \textbf{AVERAGE}&\textbf{one}&\textbf{2.570}&\textbf{2.787}&\textbf{2.617}\\\hline

        \end{tabular}
         }
        \caption{Averaged results across subject matter experts for zero-shot (zero) and one-shot (one) settings, using GPT-4o for generation. `Rel. to the evidence' refers to the `Relevance to the evidence' criterion, and `Rel. to human rec.' refers to the `Relevance to the human-created recommendation' criterion.}
        \label{humanevalReszero}
	\end{table}

 \begin{table}[hbt!]
	       \centering
        \resizebox{\linewidth}{!}{
       \begin{tabular}{|l|c|c|}\hline
          \textbf{Dataset)}&\textbf{zero-shot setting}&\textbf{one-shot setting}\\\hline\hline
            UK Care Homes&60\%&55\%\\\hline
            US Children’s Bureau&58\%&57\%\\\hline
            NSPCC reports&60\%&59\%\\\hline

        \end{tabular}
         }
            \caption{Results per dataset for criterion (5) \textit{Is the recommendation actionable?}}\label{actionable}
	\end{table}

 \paragraph{Correlation Analysis.}  
\begin{figure*}[hbt!]
		    \begin{center}
		     \includegraphics[scale = 0.47]{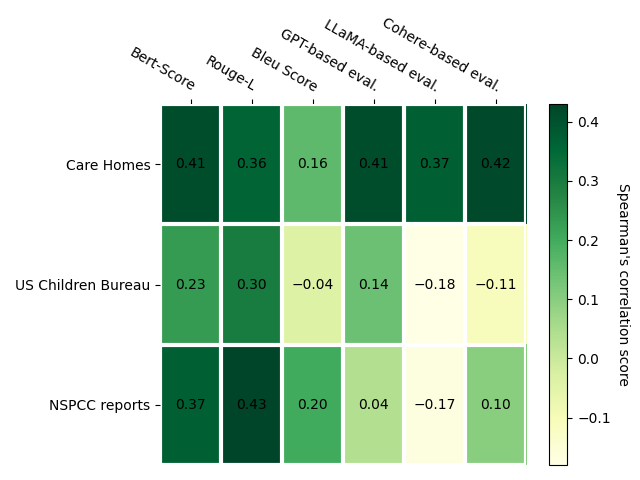}
            \includegraphics[scale = 0.47]{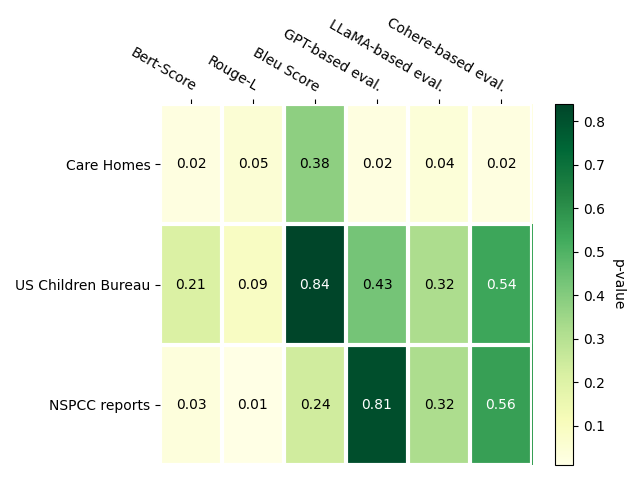}		
        \caption{Spearman's rank correlation (left) and p-values (right) between manual evaluation and automated metrics-based evaluation across the three datasets where \textit{`eval'} refers to evaluation, \textit{`Care Homes'}, \textit{`US Chidlren Bureau'} and \textit{`NSPCC reports'} refer to the results from the human-based evaluation for the Care Homes dataset, US Children Bureau, and NSPCC datasets, respectively.}\label{spearmancor_humanVSauto}
       
		    \end{center}
    \end{figure*}

  \begin{figure*}[hbt!]
		    \begin{center}
		     \includegraphics[scale = 0.44]{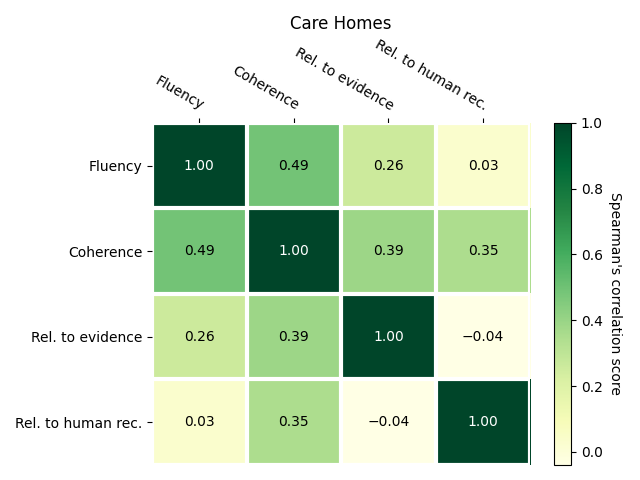}
            \includegraphics[scale = 0.44]{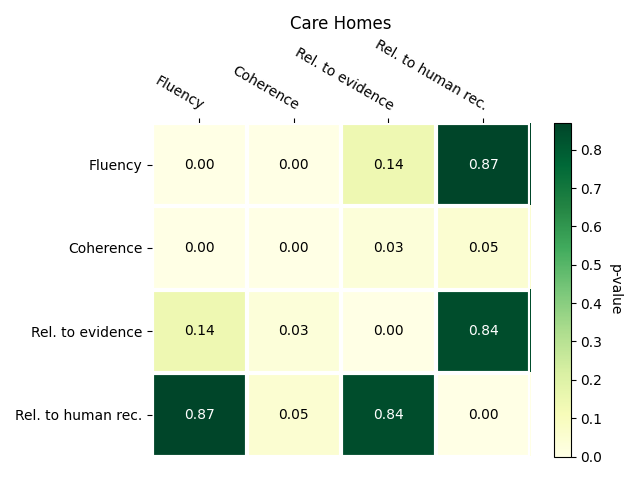}	
            \includegraphics[scale = 0.44]{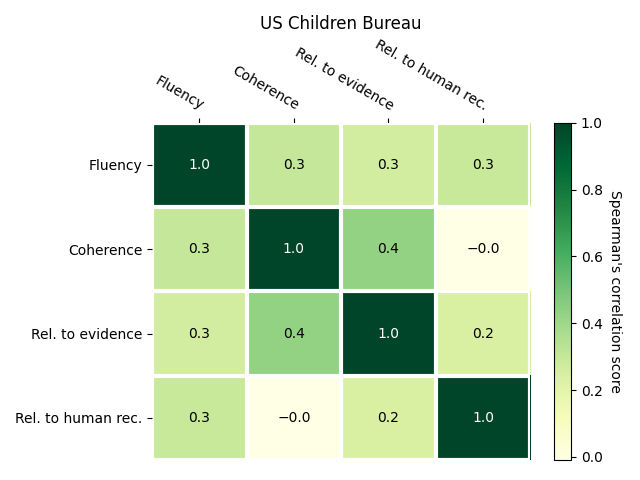}
            \includegraphics[scale = 0.44]{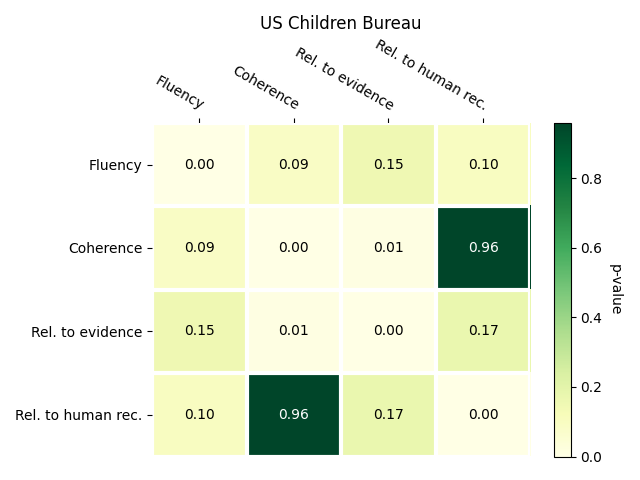}
             \includegraphics[scale = 0.44]{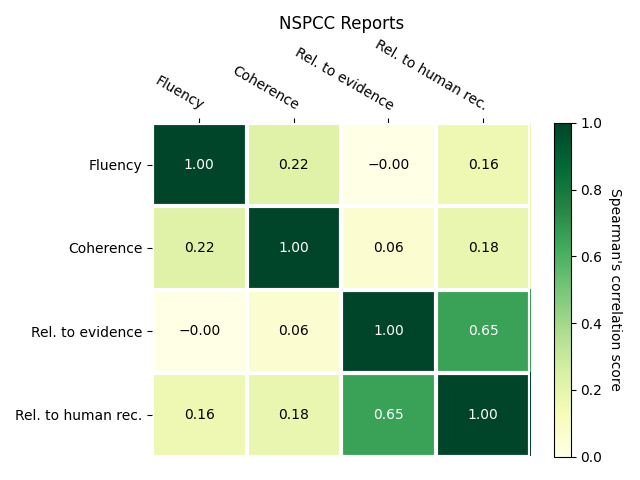}
            \includegraphics[scale = 0.44]{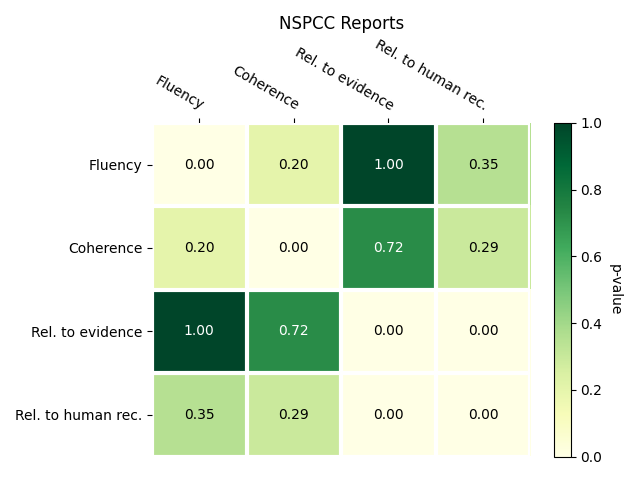}
        \caption{Spearman's rank correlation between across the criteria for the manual evaluation where \textit{`Rel. to evidence' } refers to Relevance to the evidence, \textit{`Rel. to human rec.'} reference to Relevance to the human-created recommendation.}\label{spearmancor_humancriteriacomp}
       
		    \end{center}
    \end{figure*}
 We investigated the correlation between human-based evaluation and automatic metrics considered in automatic evaluation (see Section~\ref{automaticeval}) across the three datasets. We took the average across the two annotators for each generated recommendation to compute the correlation. Figure~\ref{spearmancor_humanVSauto} shows the Spearman's rank correlation coefficient and p-value\footnote{Guidance on the Spearman's rank scale is given in the Appendix.} across the automatic metrics and the human evaluation scores regarding criteria \textit{`(4) relevance to the human-created recommendation'} (see Section~\ref{evaluationoverall}) which is the same criteria used for LLM-based evaluation. The p-values for a large proportion of the correlations are above 0.4 which makes them correlated, but not too strongly. This supports the findings in Section~\ref{reshumaneval} and suggests that the task of evaluating recommendations for these datasets is quite a complex task and requires more purpose-build metrics. Furthermore, according to the correlation analysis presented in Figure~\ref{spearmancor_humanVSauto}, no metric achieved high agreement (above 0.5) with the human annotators.  These findings highlights even further that we should not rely on a single metric to capture all quality aspects of a model's output. A surprising finding is that BERT-score and ROUGE-L tend to have better alignment with the human annotators than LLaMA and Cohere-based evaluation. However, the BLUE score shows to be the least reliable among the metrics, which is similar to findings in other NLG tasks \cite{mathur-etal-2020-tangled}. Correlation analysis across the human-based evaluation criteria per dataset (see Figure~\ref{spearmancor_humancriteriacomp}) shows that there are not significant trends of correlation relationships between the different criteria across the datasets. This suggests that despite belonging to the same domain/task, these datasets are quite diverse and require special attention of how to deal with their characteristics.


\section{Discussion}
\noindent\textbf{Potential of LLMs for the task.} Analyses using a wide range of automatic metrics and human evaluation, as presented in Section~\ref{resandanalysis}, show promising performance of LLMs on the recommendation generation task. Notably, both human and LLM-based evaluations—regardless of the model used—produced high scores, with human evaluators assigning slightly higher ratings, ranging from approximately 2.5 to 2.8 out of a maximum of 3.0. These results show the potential of state-of-the-art models to be utilised for more specialised domains to support the work of subject-matter experts. Further, the results from the human evaluation presented in Table~\ref{humanevalReszero} show a higher scoring for the \textit{`relevance to the evidence'} versus \textit{`relevance to the human-based recommendation'} criteria. This suggests that LLMs can be more suited for providing a different perspective of the problem versus simply matching the expert-created text.\\
\noindent\textbf{Hallucinations and Reliability.} A major challenge in LLM-based text generation is the risk of hallucinations~\cite{ji2023survey,filippova2020controlled}, with solutions varying depending on the task and available resources. We note that addressing this issue is beyond the scope of this paper. However, our human-based evaluation approach helps identify discrepancies within the dataset. For example, Criterion (3) from the evaluation framework (see Section~\ref{evaluationoverall}) assesses whether a generated recommendation is meaningfully related to the given evidence. The average score for this criterion exceeds 2.5 (on a 3-point scale) across all datasets, indicating strong relevance. Additionally, subject matter experts found a large proportion of the recommendations to be practically applicable to policy-making processes (Criterion (5), Section~\ref{evaluationoverall}). These findings suggest a minimal presence of hallucinations in the generated content. Nonetheless, we believe that future research should include more rigorous analysis and the development of evaluation methods that can ensure higher dataset reliability.\\
\noindent\textbf{Evaluation metrics for text generation.}
A comparison between the different automated metrics (see Section~\ref{automaticeval}) and the correlation analysis between automated and human-based evaluation (Section~\ref{reshumaneval}) highlights the limitations of traditional evaluation metrics such as BLEU for more complex NLG tasks such as recommendation generation.  Further, LLM-based metrics and human-based evaluation showed similar satisfactory results suggesting good performance of text generation models for the given task. However, correlation analysis showed that no metric achieved high agreement with the human evaluators which suggests that when it comes to complex NLG tasks, we should not rely on a single metric. The relatively low scores from the inter-annotator agreement analysis illustrates further the complexity of the task, which proved challenging even for domain experts. In future work, the human evaluation can be expanded to include additional criteria, such as level of detail, feasibility within specific institutional settings, or sensitivity to policy context. However, we believe that the criteria used in this study reflect the most important aspects of recommendation quality in this domain. Fluency and coherence assess clarity and structure; relevance to the evidence ensures grounding in the provided information; relevance to the human-created recommendation allows comparison with expert judgement; and actionability captures whether the recommendation can realistically inform policy and agency work.
In future, the study can also be expanded by incorporating more qualitative studies and more purpose-oriented metrics tailored to recommendation generation in applied policy settings.

\section{Conclusions}
This paper introduces the first comprehensive effort to leverage LLMs for the specialized NLG task of recommendation generation aimed at informing policy decisions and enhancing the work of public service agencies. We release a unified benchmark dataset for this task, PubRec-Bench, compiled from three distinct data sources. We evaluate three state-of-the-art models at the time we performed our experiments, namely GPT-4o, Cohere’s Command R+, and LLaMA 3, using both LLM-based and human evaluations, yielding promising results. Human evaluators judged most generated recommendations as highly relevant to the provided evidence and consistently coherent and fluent in both structure and content. Additionally, subject matter experts rated the majority of outputs as actionable, meaning they offer practical, real-world utility. Finally, we provide a thorough analysis of evaluation methodologies, highlighting the need for more task- and purpose-specific metrics tailored to the demands of NLG in applied, real-world settings.
\newpage

\section*{Limitations}

This study represents a first step towards using LLMs for recommendation generation to support policy making and agency work, and it comes with several limitations. 

First, the datasets are available in English only and are drawn from UK and US sources. While combining these contexts introduces variation in institutional and legal frameworks, which strengthens the diversity of the benchmark, both settings remain English-speaking Western policy environments. Differences in governance structures and reporting conventions may influence how recommendations are formulated, and the findings may therefore not generalise to non-English or non-Western contexts where policy-making practices differ.

Second, our analyses are conducted primarily in zero-shot and one-shot settings. While this reflects realistic deployment scenarios, further work is needed to explore how performance may change under alternative adaptation strategies or more domain-specific tuning approaches.

Third, the corpus consists of three datasets of relatively limited size. Although the reports cover sensitive and high-impact domains, the number of available documents is constrained by the nature of serious case reviews and inspection processes. Future work should aim to expand the dataset by incorporating reports from additional sources and policy areas.

Finally, as all three datasets are sourced from publicly available websites, it is possible that portions of these reports were included in the pre-training data of some large language models. Due to the proprietary nature of training corpora, this cannot be verified directly. Although the reports represent a small and specialised subset of public documents, the possibility of partial overlap cannot be fully excluded. Future studies could address this by evaluating models on newly released or temporally held-out reports.

 \section*{Ethical Considerations}
 The goal of our method is to facilitate rather than replace practitioners in the public sector in writing high quality recommendations. Specifically, we hope that LLM-generated recommendations can provide a different and useful aspect of the problem at hand and also facilitate more efficient decision-making for practitioners. Given the domain at hand, we believe that subject matter experts should take pivotal role in recommendation creation, but it is also important to find ways to utilise LLMs strengths to support the work of experts.

\nocite{*}
\section{Bibliographical References}\label{sec:reference}

\bibliographystyle{lrec2026-natbib}
\bibliography{lrec2026-example}

@inproceedings{mathur-etal-2020-tangled,
    title = "Tangled up in {BLEU}: Reevaluating the Evaluation of Automatic Machine Translation Evaluation Metrics",
    author = "Mathur, Nitika  and
      Baldwin, Timothy  and
      Cohn, Trevor",
    editor = "Jurafsky, Dan  and
      Chai, Joyce  and
      Schluter, Natalie  and
      Tetreault, Joel",
    booktitle = "Proceedings of the 58th Annual Meeting of the Association for Computational Linguistics",
    month = jul,
    year = "2020",
    address = "Online",
    publisher = "Association for Computational Linguistics",
    url = "https://aclanthology.org/2020.acl-main.448/",
    doi = "10.18653/v1/2020.acl-main.448",
    pages = "4984--4997"
}

@inproceedings{kocmi2023large,
  title={Large Language Models Are State-of-the-Art Evaluators of Translation Quality},
  author={Kocmi, Tom and Federmann, Christian},
  booktitle={Proceedings of the 24th Annual Conference of the European Association for Machine Translation},
  pages={193--203},
  year={2023}
}

@article{bishop2023longdocfactscore,
  title={LongDocFACTScore: Evaluating the Factuality of Long Document Abstractive Summarisation},
  author={Bishop, Jennifer and Xie, Qianqian and Ananiadou, Sophia},
  journal={CoRR},
  year={2023}
}

@inproceedings{li2024enhancing,
  title={Enhancing Court View Generation with Knowledge Injection and Guidance},
  author={Li, Ang and Wu, Yiquan and Liu, Yifei and Kuang, Kun and Wu, Fei and Cai, Ming},
  booktitle={Proceedings of the 2024 Joint International Conference on Computational Linguistics, Language Resources and Evaluation (LREC-COLING 2024)},
  pages={5896--5906},
  year={2024}
}

@inproceedings{chaganty-etal-2018-price,
    title = "The price of debiasing automatic metrics in natural language evalaution",
    author = "Chaganty, Arun  and
      Mussmann, Stephen  and
      Liang, Percy",
    editor = "Gurevych, Iryna  and
      Miyao, Yusuke",
    booktitle = "Proceedings of the 56th Annual Meeting of the Association for Computational Linguistics (Volume 1: Long Papers)",
    month = jul,
    year = "2018",
    address = "Melbourne, Australia",
    publisher = "Association for Computational Linguistics",
    url = "https://aclanthology.org/P18-1060",
    doi = "10.18653/v1/P18-1060",
    pages = "643--653",
}

@article{khashabi2021genie,
  title={GENIE: Toward Reproducible and Standardized Human Evaluation for Text Generation},
  author={Khashabi, Daniel and Stanovsky, Gabriel and Bragg, Jonathan and Lourie, Nicholas and Kasai, Jungo and Choi, Yejin and Smith, Noah A and Weld, Daniel S},
  journal={Proceedings of the 2022 Conference on Empirical Methods in Natural Language Processing},
  pages={11444--11458},
  year={2022}
}

@article{gao2024llm,
  title={Llm-based nlg evaluation: Current status and challenges},
  author={Gao, Mingqi and Hu, Xinyu and Yin, Xunjian and Ruan, Jie and Pu, Xiao and Wan, Xiaojun},
  journal={Computational Linguistics},
  pages={1--28},
  year={2025},
  publisher={MIT Press 255 Main Street, 9th Floor, Cambridge, Massachusetts 02142, USA~…}
}

@inproceedings{10.1145/3640544.3645216, author = {Desmond, Michael and Ashktorab, Zahra and Pan, Qian and Dugan, Casey and Johnson, James M.}, title = {EvaluLLM: LLM assisted evaluation of generative outputs}, year = {2024}, isbn = {9798400705090}, publisher = {Association for Computing Machinery}, address = {New York, NY, USA}, url = {https://doi.org/10.1145/3640544.3645216}, doi = {10.1145/3640544.3645216}, booktitle = {Companion Proceedings of the 29th International Conference on Intelligent User Interfaces}, pages = {30–32}, numpages = {3}, location = {, Greenville, SC, USA, }, series = {IUI '24 Companion} }

@inproceedings{liusie-etal-2024-llm,
    title = "{LLM} Comparative Assessment: Zero-shot {NLG} Evaluation through Pairwise Comparisons using Large Language Models",
    author = "Liusie, Adian  and
      Manakul, Potsawee  and
      Gales, Mark",
    editor = "Graham, Yvette  and
      Purver, Matthew",
    booktitle = "Proceedings of the 18th Conference of the European Chapter of the Association for Computational Linguistics (Volume 1: Long Papers)",
    month = mar,
    year = "2024",
    address = "St. Julian{'}s, Malta",
    publisher = "Association for Computational Linguistics",
    url = "https://aclanthology.org/2024.eacl-long.8",
    pages = "139--151",
}

@article{panickssery2024llm,
 title={Llm evaluators recognize and favor their own generations},
  author={Panickssery, Arjun and Bowman, Samuel and Feng, Shi},
  journal={Advances in Neural Information Processing Systems},
  volume={37},
  pages={68772--68802},
  year={2024}
}

@article{sheng2024repeval,
  title={RepEval: Effective Text Evaluation with LLM Representation},
  author={Sheng, Shuqian and Xu, Yi and Zhang, Tianhang and Shen, Zanwei and Fu, Luoyi and Ding, Jiaxin and Zhou, Lei and Gan, Xiaoying and Wang, Xinbing and Zhou, Chenghu},
  journal={Proceedings of the 2024 Conference on Empirical Methods in Natural Language Processing},
  pages={7019--7033},
  year={2024}
}

@article{shi2022language,
  title={Language models are multilingual chain-of-thought reasoners},
  author={Shi, Freda and Suzgun, Mirac and Freitag, Markus and Wang, Xuezhi and Srivats, Suraj and Vosoughi, Soroush and Chung, Hyung Won and Tay, Yi and Ruder, Sebastian and Zhou, Denny and others},
  journal={The Eleventh International Conference on Learning Representations},
  year={2022}
}

@inproceedings{reynolds2021prompt,
  title={Prompt programming for large language models: Beyond the few-shot paradigm},
  author={Reynolds, Laria and McDonell, Kyle},
  booktitle={Extended Abstracts of the 2021 CHI Conference on Human Factors in Computing Systems},
  pages={1--7},
  year={2021}
}

@inproceedings{lin2004rouge,
  title={Rouge: A package for automatic evaluation of summaries},
  author={Lin, Chin-Yew},
  booktitle={Text summarization branches out},
  pages={74--81},
  year={2004}
}

@article{chhun2022human,
  title={Of Human Criteria and Automatic Metrics: A Benchmark of the Evaluation of Story Generation},
  author={Chhun, Cyril and Colombo, Pierre and Suchanek, Fabian M and Clavel, Chlo{\'e}},
  journal={29th International Conference on Computational Linguistics (COLING 2022)},
  year={2022}
}

@article{brown2020language,
  title={Language models are few-shot learners},
  author={Brown, Tom and Mann, Benjamin and Ryder, Nick and Subbiah, Melanie and Kaplan, Jared D and Dhariwal, Prafulla and Neelakantan, Arvind and Shyam, Pranav and Sastry, Girish and Askell, Amanda and others},
  journal={Advances in neural information processing systems},
  volume={33},
  pages={1877--1901},
  year={2020}
}

@article{schick2020s,
  title={It’s Not Just Size That Matters: Small Language Models Are Also Few-Shot Learners},
  author={Schick, Timo and Sch{\"u}tze, Hinrich},
  journal={Proceedings of the 2021 Conference of the North American Chapter of the Association for Computational Linguistics: Human Language Technologies},
  pages={2339--2352},
  year={2021}
}

@article{tang2022context,
   title={Context-Tuning: Learning Contextualized Prompts for Natural Language Generation},
  author={Tang, Tianyi and Li, Junyi and Zhao, Wayne Xin and Wen, Ji-Rong},
  journal={Proceedings of the 29th International Conference on Computational Linguistics},
  pages={6340--6354},
  year={2022}
}

@article{labrak2023zero,
  title={A Zero-shot and Few-shot Study of Instruction-Finetuned Large Language Models Applied to Clinical and Biomedical Tasks},
  author={Labrak, Yanis and Rouvier, Micka{\"e}l and Dufour, Richard},
  journal={Fourteenth Language Resources and Evaluation Conference (LREC-COLING 2024)},
  year={2024}
}

@inproceedings{gao2021making,
  title={Making Pre-trained Language Models Better Few-shot Learners},
  author={Gao, Tianyu and Fisch, Adam and Chen, Danqi},
  journal={Proceedings of the 59th Annual Meeting of the Association for Computational Linguistics and the 11th International Joint Conference on Natural Language Processing (Volume 1: Long Papers)},
  pages={3816--3830},
  year={2021}
}

@article{mosbach2023few,
  title={Few-shot Fine-tuning vs. In-context Learning: A Fair Comparison and Evaluation},
  author={Mosbach, Marius and Pimentel, Tiago and Ravfogel, Shauli and Klakow, Dietrich and Elazar, Yanai},
  journal={Findings of the Association for Computational Linguistics: ACL 2023},
  pages={12284--12314},
  year={2023}
}

@inproceedings{schick2021s,
  title={It’s Not Just Size That Matters: Small Language Models Are Also Few-Shot Learners},
  author={Schick, Timo and Sch{\"u}tze, Hinrich},
  booktitle={Proceedings of the 2021 Conference of the North American Chapter of the Association for Computational Linguistics: Human Language Technologies},
  pages={2339--2352},
  year={2021}
}

@inproceedings{le2021many,
  title={How many data points is a prompt worth?},
  author={Le Scao, Teven and Rush, Alexander M},
  booktitle={Proceedings of the 2021 Conference of the North American Chapter of the Association for Computational Linguistics: Human Language Technologies},
  pages={2627--2636},
  year={2021}
}

@article{sun2023evaluating,
   title={Evaluating the Zero-shot Robustness of Instruction-tuned Language Models},
  author={Sun, Jiuding and Shaib, Chantal and Wallace, Byron C},
  journal={The Twelfth International Conference on Learning Representations},
  year={2023}
}

@article{wang2022automatic,
  title={Automatic Multi-Label Prompting: Simple and Interpretable Few-Shot Classification},
  author={Wang, Han and Xu, Canwen and McAuley, Julian},
  booktitle={Proceedings of the 2022 Conference of the North American Chapter of the Association for Computational Linguistics: Human Language Technologies},
  pages={5483--5492},
  year={2022}
}

@article{shin2020autoprompt,
  title={AutoPrompt: Eliciting Knowledge from Language Models with Automatically Generated Prompts},
  author={Shin, Taylor and Razeghi, Yasaman and Logan IV, Robert L and Wallace, Eric and Singh, Sameer},
  booktitle={Proceedings of the 2020 Conference on Empirical Methods in Natural Language Processing (EMNLP)},
  pages={4222--4235},
  year={2020}
}

@inproceedings{zhang-etal-2022-learn,
    title = "Learn to Adapt for Generalized Zero-Shot Text Classification",
    author = "Zhang, Yiwen  and
      Yuan, Caixia  and
      Wang, Xiaojie  and
      Bai, Ziwei  and
      Liu, Yongbin",
    booktitle = "Proceedings of the 60th Annual Meeting of the Association for Computational Linguistics (Volume 1: Long Papers)",
    month = may,
    year = "2022",
    address = "Dublin, Ireland",
    publisher = "Association for Computational Linguistics",
    url = "https://aclanthology.org/2022.acl-long.39",
    doi = "10.18653/v1/2022.acl-long.39",
    pages = "517--527",
   
}

@inproceedings{schonfeld2019generalized,
  title={Generalized Zero-and Few-Shot Learning via Aligned Variational Autoencoders},
  author={Sch{\"o}nfeld, Edgar and Ebrahimi, Sayna and Sinha, Samarth and Darrell, Trevor and Akata, Zeynep},
  booktitle={2019 IEEE/CVF Conference on Computer Vision and Pattern Recognition (CVPR)},
  pages={8239--8247},
  year={2019},
  organization={IEEE}
}

@inproceedings{song2021generalized,
  title={Generalized zero-shot text classification for ICD coding},
  author={Song, Congzheng and Zhang, Shanghang and Sadoughi, Najmeh and Xie, Pengtao and Xing, Eric},
  booktitle={Proceedings of the Twenty-Ninth International Conference on International Joint Conferences on Artificial Intelligence},
  pages={4018--4024},
  year={2021}
}

@article{oniani2023large,
  title={Large Language Models Vote: Prompting for Rare Disease Identification},
  author={Oniani, David and Hilsman, Jordan and Dong, Hang and Gao, Fengyi and Verma, Shiven and Wang, Yanshan},
  journal={arXiv preprint arXiv:2308.12890},
  year={2023}
}

@inproceedings{papineni2002bleu,
  title={Bleu: a method for automatic evaluation of machine translation},
  author={Papineni, Kishore and Roukos, Salim and Ward, Todd and Zhu, Wei-Jing},
  booktitle={Proceedings of the 40th annual meeting of the Association for Computational Linguistics},
  pages={311--318},
  year={2002}
}

@misc{safeguard,
    title = {Safeguarding people},
    url = { https://www.cqc. org.uk/what-we-do/how-we-do-our-job/safeguarding-people},
    author = {Care Quality Commission.},
    year = {2014},
    note = {07, 2024}
}

@article{zhang2019bertscore,
  title={BERTScore: Evaluating Text Generation with BERT},
  author={Zhang, Tianyi and Kishore, Varsha and Wu, Felix and Weinberger, Kilian Q and Artzi, Yoav},
  journal={International Conference on Learning Representations},
  year={2019}
}

@inproceedings{yue2021circumstances,
  title={Circumstances enhanced criminal court view generation},
  author={Yue, Linan and Liu, Qi and Wu, Han and An, Yanqing and Wang, Li and Yuan, Senchao and Wu, Dayong},
  booktitle={Proceedings of the 44th International ACM SIGIR Conference on Research and Development in Information Retrieval},
  pages={1855--1859},
  year={2021}
}

@article{yu2022legal,
  title={Legal prompting: Teaching a language model to think like a lawyer},
  author={Yu, Fangyi and Quartey, Lee and Schilder, Frank},
  journal={arXiv preprint arXiv:2212.01326},
  year={2022}
}

@article{zhang2024benchmarking,
  title={Benchmarking large language models for news summarization},
  author={Zhang, Tianyi and Ladhak, Faisal and Durmus, Esin and Liang, Percy and McKeown, Kathleen and Hashimoto, Tatsunori B},
  journal={Transactions of the Association for Computational Linguistics},
  volume={12},
  pages={39--57},
  year={2024},
  publisher={MIT Press One Broadway, 12th Floor, Cambridge, Massachusetts 02142, USA~…}
}

@article{xie2023survey,
  title={A survey for biomedical text summarization: From pre-trained to large language models},
  author={Xie, Qianqian and Luo, Zhehengz and Wang, Benyou and Ananiadou, Sophia},
  journal={arXiv preprint arXiv:2304.08763},
  year={2023}
}

@article{viswanathan2023prompt2model,
  title={Prompt2Model: Generating Deployable Models from Natural Language Instructions},
  author={Viswanathan, Vijay and Zhao, Chenyang and Bertsch, Amanda and Wu, Tongshuang and Neubig, Graham},
  booktitle={Proceedings of the 2023 Conference on Empirical Methods in Natural Language Processing: System Demonstrations},
  pages={413--421},
  year={2023}
}

@article{plaza2023leveraging,
  title={Large language models for text classification: From zero-shot learning to fine-tuning},
  author={Chae, Youngjin and Davidson, Thomas},
  journal={Open Science Foundation},
  volume={10},
  year={2023}
}

@article{huang2024good,
  title={How Good Are Low-bit Quantized LLaMA3 Models? An Empirical Study},
  author={Huang, Wei and Ma, Xudong and Qin, Haotong and Zheng, Xingyu and Lv, Chengtao and Chen, Hong and Luo, Jie and Qi, Xiaojuan and Liu, Xianglong and Magno, Michele},
  journal={CoRR},
  year={2024}
}

@article{wu2023precedent,
  title={Precedent-Enhanced Legal Judgment Prediction with LLM and Domain-Model Collaboration},
  author={Wu, Yiquan and Zhou, Siying and Liu, Yifei and Lu, Weiming and Liu, Xiaozhong and Zhang, Yating and Sun, Changlong and Wu, Fei and Kuang, Kun},
  journal={Proceedings of the 2023 Conference on Empirical Methods in Natural Language Processing},
  pages={12060--12075},
  year={2023}
}

@article{spearman1961proof,
  title={The proof and measurement of association between two things.},
  author={Spearman, Charles},
  year={1961},
  publisher={Appleton-Century-Crofts}
}

@article{landis1977measurement,
  title={The measurement of observer agreement for categorical data},
  author={Landis, J Richard and Koch, Gary G},
  journal={biometrics},
  pages={159--174},
  year={1977},
  publisher={JSTOR}
}

@article{katz2023natural,
  title={Natural Language Processing in the Legal Domain},
  author={Katz, Daniel Martin and Hartung, Dirk and Gerlach, Lauritz and Jana, Abhik and Bommarito, Michael James},
  journal={Available at SSRN 4336224},
  year={2023}
}

@article{li2024pre,
  title={Pre-trained language models for text generation: A survey},
  author={Li, Junyi and Tang, Tianyi and Zhao, Wayne Xin and Nie, Jian-Yun and Wen, Ji-Rong},
  journal={ACM Computing Surveys},
  volume={56},
  number={9},
  pages={1--39},
  year={2024},
  publisher={ACM New York, NY}
}

@article{li2019generating,
  title={Generating Long and Informative Reviews with Aspect-Aware Coarse-to-Fine Decoding},
  author={Li, Junyi and Zhao, Wayne Xin and Wen, Ji-Rong and Song, Yang},
  booktitle={Proceedings of the 57th Annual Meeting of the Association for Computational Linguistics},
  pages={1969--1979},
  year={2019}
}

@article{yang2022re3,
  title={Re3: Generating Longer Stories With Recursive Reprompting and Revision},
  author={Yang, Kevin and Tian, Yuandong and Peng, Nanyun and Klein, Dan},
  booktitle={Proceedings of the 2022 Conference on Empirical Methods in Natural Language Processing},
  pages={4393--4479},
  year={2022}
}

@article{xie2023knowledge,
  title={Knowledge-enhanced graph topic transformer for explainable biomedical text summarization},
  author={Xie, Qianqian and Tiwari, Prayag and Ananiadou, Sophia},
  journal={IEEE journal of biomedical and health informatics},
  year={2023},
  publisher={IEEE}
}

@article{lopez2024evaluation,
  title={Evaluation of large language models performance against humans for summarizing MRI knee radiology reports: A feasibility study},
  author={L{\'o}pez-{\'U}beda, Pilar and Mart{\'\i}n-Noguerol, Teodoro and D{\'\i}az-Angulo, Carolina and Luna, Antonio},
  journal={International Journal of Medical Informatics},
  volume={187},
  pages={105443},
  year={2024},
  publisher={Elsevier}
}

@article{savelka2023can,
  title={Can gpt-4 support analysis of textual data in tasks requiring highly specialized domain expertise?},
  author={Savelka, Jaromir and Ashley, Kevin D and Gray, Morgan A and Westermann, Hannes and Xu, Huihui},
  journal={arXiv preprint arXiv:2306.13906},
  year={2023}
}

@article{filippova2020controlled,
  title={Controlled Hallucinations: Learning to Generate Faithfully from Noisy Data},
  author={Filippova, Katja},
  journal={Findings of the Association for Computational Linguistics: EMNLP 2020},
  pages={864--870},
  year={2020}
}

@article{ji2023survey,
  title={Survey of hallucination in natural language generation},
  author={Ji, Ziwei and Lee, Nayeon and Frieske, Rita and Yu, Tiezheng and Su, Dan and Xu, Yan and Ishii, Etsuko and Bang, Ye Jin and Madotto, Andrea and Fung, Pascale},
  journal={ACM Computing Surveys},
  volume={55},
  number={12},
  pages={1--38},
  year={2023},
  publisher={ACM New York, NY}
}

@article{goanta2023regulation,
  title={Regulation and NLP (RegNLP): Taming Large Language Models},
  author={Goanț{\u{a}}, C{\u{a}}t{\u{a}}lina and Aletras, Nikolaos and Chalkidis, Ilias and Ranchord{\'a}s, Sofia and Spanakis, Gerasimos},
  journale={Proceedings of the 2023 Conference on Empirical Methods in Natural Language Processing},
  pages={8712--8724},
  year={2023}
}

@article{touvron2023llama2,
  title={Llama 2: Open Foundation and Fine-Tuned Chat Models},
  author={Touvron, Hugo and Martin, Louis and Stone, Kevin and Albert, Peter and Almahairi, Amjad and Babaei, Yasmine and Bashlykov, Nikolay and Batra, Soumya and Bhargava, Prajjwal and Bhosale, Shruti and others},
  booktitle={arXiv e-prints},
publisher={arXiv},
address={online},
  pages={arXiv--2307},
  year={2023}
}

@article{dubey2024llama,
  title={The llama 3 herd of models},
  author={Dubey, Abhimanyu and Jauhri, Abhinav and Pandey, Abhinav and Kadian, Abhishek and Al-Dahle, Ahmad and Letman, Aiesha and Mathur, Akhil and Schelten, Alan and Yang, Amy and Fan, Angela and others},
  journal={arXiv preprint arXiv:2407.21783},
  year={2024}
}

@article{EMILLE,
    author = "{Anthony McEnery and others}",
    title = "The EMILLE/CIIL Corpus",
    publisher = "distributed via ELRA: ELRA-Id W0037",
    year = "2004",
    islrn = "039-846-040-604-0",
    organization = "EMILLE (Enabling Minority Language Engineering) Project"
}

@article{razumovskaia-etal-2024-little-red,
    title = "Little Red Riding Hood Goes around the Globe: Crosslingual Story Planning and Generation with Large Language Models",
    author = "Razumovskaia, Evgeniia  and
      Maynez, Joshua  and
      Louis, Annie  and
      Lapata, Mirella  and
      Narayan, Shashi",
    editor = "Calzolari, Nicoletta  and
      Kan, Min-Yen  and
      Hoste, Veronique  and
      Lenci, Alessandro  and
      Sakti, Sakriani  and
      Xue, Nianwen",
    booktitle = "Proceedings of the 2024 Joint International Conference on Computational Linguistics, Language Resources and Evaluation (LREC-COLING 2024)",
    month = may,
    year = "2024",
    address = "Torino, Italia",
    publisher = "ELRA and ICCL",
    url = "https://aclanthology.org/2024.lrec-main.929",
    pages = "10616--10631",
}

@article{chowdhery2023palm,
  title={Palm: Scaling language modeling with pathways},
  author={Chowdhery, Aakanksha and Narang, Sharan and Devlin, Jacob and Bosma, Maarten and Mishra, Gaurav and Roberts, Adam and Barham, Paul and Chung, Hyung Won and Sutton, Charles and Gehrmann, Sebastian and others},
  journal={Journal of Machine Learning Research},
  volume={24},
  number={240},
  pages={1--113},
  year={2023}
}

@article{agrawal2023qameleon,
  title={Qameleon: Multilingual qa with only 5 examples},
  author={Agrawal, Priyanka and Alberti, Chris and Huot, Fantine and Maynez, Joshua and Ma, Ji and Ruder, Sebastian and Ganchev, Kuzman and Das, Dipanjan and Lapata, Mirella},
  journal={Transactions of the Association for Computational Linguistics},
  volume={11},
  pages={1754--1771},
  year={2023},
  publisher={MIT Press One Broadway, 12th Floor, Cambridge, Massachusetts 02142, USA~…}
}

@article{tyss2024lexabsumm,
  title={LexAbSumm: Aspect-based Summarization of Legal Decisions},
  author={Tyss, Santosh and Aly, Mahmoud and Grabmair, Matthias},
  booktitle={Proceedings of the 2024 Joint International Conference on Computational Linguistics, Language Resources and Evaluation (LREC-COLING 2024)},
  pages={10422--10431},
  year={2024}
}

@article{wolf2019huggingface,
  title={Huggingface's transformers: State-of-the-art natural language processing},
  author={Wolf, Thomas and Debut, Lysandre and Sanh, Victor and Chaumond, Julien and Delangue, Clement and Moi, Anthony and Cistac, Pierric and Rault, Tim and Louf, R{\'e}mi and Funtowicz, Morgan and others},
  journal={arXiv preprint arXiv:1910.03771},
  year={2019}
}

\section{Language Resource References}
\label{lr:ref}
\bibliographystylelanguageresource{lrec2026-natbib}
\bibliographylanguageresource{languageresource}
\appendix

\section{Appendix}\label{sec:appendix}

\subsection{Model parameters and Computational Budget}\label{modelparam}
The model parameters we used for generating recommendations are as follows: (1) For GPT4-o and Cohere-based model we have used temperature of 0.7 and for LLaMA a temperature of 0.6. These are the default values recommended for these models. We used 7 hours of GPU budget and Nvidia RTX 4090 GPU.

\subsection{Human-Based Evaluation}\label{humaneval}

\begin{figure}[hbt!]
		    \begin{center}
		     \includegraphics[scale = 0.3]{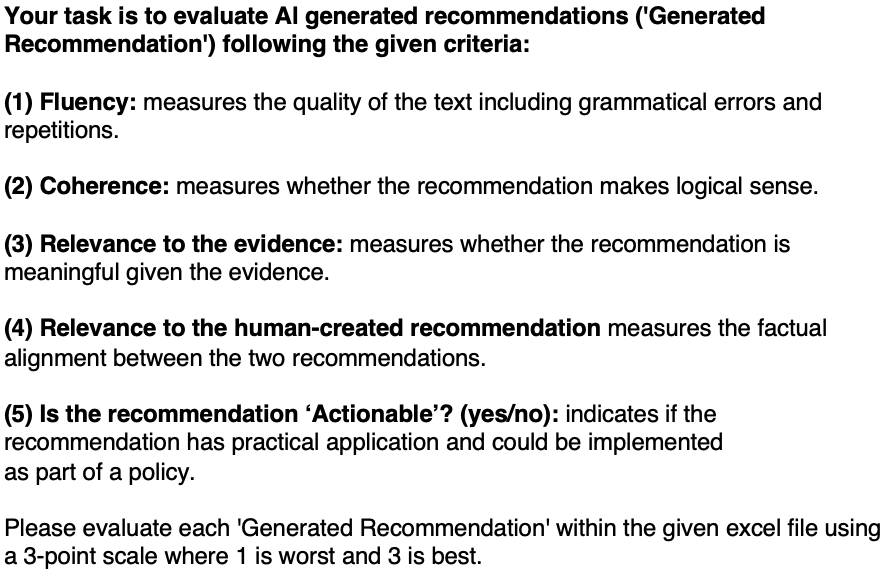}
		
        \caption{Instructions for human evaluation.}\label{instruct}
       
		    \end{center}
    \end{figure}

Figure~\ref{instruct} shows the instructions given to the annotators in order to perform the human-based evaluation. The five annotators were selected through an interview process based on their subject-matter expertise. They were compensated at an hourly rate aligned with standard payment guidelines, as approved by JobShop and University guidelines.

 \subsection{Spearman's Rank Correlation}\label{spearman}
 The Spearman's Rank Correlation Coefficient is a statistical measure of the strength of the relationship between two sets of data. A description of the strength of correlation is given in Table~\ref{spearmanguidence}. The p-value  is the probability of how likely it is that any observed correlation is due to chance. A p-value close to 1 suggests no correlation other than due to chance. If your p-value is close to 0, the observed correlation is unlikely to be due to chance.

  \begin{table}[hbt!]
	       \centering
        \resizebox{\linewidth}{!}{
       \begin{tabular}{c||c}\hline
          \textbf{Value of coefficient (pos. or neg.)}&\textbf{Meaning}\\\hline\hline
            0.00-0.19&A very weak correlation\\\hline
            0.20-0.39&A weak correlation\\\hline
            0.40-0.69&A moderate correlation\\\hline
            0.70-0.89&A strong correlation\\\hline
            0.90-1.00&A very strong correlation\\\hline
  
        \end{tabular}
         }
            \caption{Interpretation of the Spearman's correlation coefficient.}\label{spearmanguidence}
	\end{table}

       

The p-value for the majority of criteria is less than 0.05 which makes the majority of correlations statistically significant. Figure~\ref{spearmancor_humancriteriacomp} shows that there are not significant trends of correlation relationships between the different criteria across the datasets. This suggests that despite belonging to the same domain/task, these datasets are quite diverse and require special attention of how to deal with their characteristics. 

\end{document}